\documentclass[aps,prl,preprint,groupedaddress]{revtex4-1}
 \usepackage{graphicx}
 \usepackage{dcolumn}
 \usepackage{bm}
 \usepackage{amsmath}
 \usepackage{color}
 \usepackage{amssymb}
 \usepackage{mathrsfs}
 \usepackage{MnSymbol}%
 \usepackage{wasysym}%
 \usepackage{gensymb}
 \usepackage{framed,color}
 \usepackage{enumitem}

\begin{document}


\title{Spin-triplet pairing state evidenced by half-quantum flux in a noncentrosymmetric superconductor}


\author{Xiaoying Xu, Yufan Li, C. -L. Chien}
\affiliation{Department of Physics and Astronomy, The Johns Hopkins University, Baltimore, MD 21218, USA}


\date{\today}

\begin{abstract}
	A prime category of superconducting materials in which to look for spin-triplet pairing and topological superconductivity are superconductors without inversion symmetry. 
	It is predicted that the breaking of parity symmetry gives rise to an admixture of spin-singlet / spin-triplet pairing states; a triplet pairing component, being substantial, seems all but guaranteed. However, the experimental confirmation of pair mixing in any particular material remains elusive. In this work, we perform phase-sensitive experiment to examine the pairing state of noncentrosymmetric superconductor $\alpha-$BiPd. The Little-Parks effect observed in mesoscopic polycrystalline $\alpha-$BiPd ring devices reveals the presence of half-integer magnetic flux quantization, which provides a decisive evidence for the spin-triplet pairing state. We find mixed half-quantum fluxes and integer-quantum fluxes, consistent with the scenario of singlet-triplet pair mixing. 

\end{abstract}

\pacs{}

\maketitle

\begin{text}
	
A superconducting Cooper pair is a system of two spin-$\frac{1}{2}$ particles, whose spin angular momentum is either 0 as a spin-singlet state, or 1 as a spin-triplet state. 
The spin-singlet pairing is found to be the case for the overwhelming majority of known superconductors (SCs), including the conventional $s$-wave SCs and the $d$-wave high-$T_c$ cuprates. 
In contrast, there are far fewer solid-state superconducting materials that manifest spin-triplet pairing. 
The efforts to search for spin-triplet SCs intensified in recent years with the surging interest in topological superconductors \cite{qi_topological_2011}.
It is shown that with few exceptions, spin-triplet SCs are inherently topological \cite{Fu2010,Sato2010,Qi2010,sato_topological_2017} and therefore ideal systems for realizing Majorana fermions \cite{read_paired_2000,kitaev_unpaired_2001}.

For SCs with inversion symmetry, the parity symmetry imposes constraint on the pairing state, which must be either spin-singlet for even-parity or spin-triplet for odd-parity \cite{anderson_structure_1984}. 
For noncentrosymmetric SCs, on the other hand, the broken parity symmetry shall give rise to an admixture of singlet and triplet pairing states \cite{gorkov_superconducting_2001,frigeri_superconductivity_2004,Fujimoto2007,yip_noncentrosymmetric_2014}, which presumably applies to all materials in this category.  
Although there is no short supply of superconducting materials without inversion symmetry, many are nonetheless behaving as conventional $s$-wave SCs \cite{yip_noncentrosymmetric_2014,Bauer2012}. 
Among them is monoclinic $\alpha-$BiPd, the first superconductor identified with noncentrosymmetric crystal structure (space group $P2_1$) \cite{alekseevskii1952sverkhprovodimost,Bhatt1979}. 
Experimental results from scanning tunneling spectroscopy \cite{sun_dirac_2015}, upper critical field and heat capacity measurements \cite{joshi_superconductivity_2011,Peets2016} indicate that the superconducting state is predominately $s$-wave with a nodeless single gap. 
This has lead to the thinking that the parity-breaking spin-orbit coupling induced by noncentrosymmetry may be too weak to realize any observable effect \cite{Peets2016,Bauer2012,sato_topological_2017}. 
However, this verdict and some of the results are at odd with the findings of multiple superconducting gaps observed by point-contact Andreev reflection \cite{mondal_andreev_2012} and penetration depth measurement \cite{Jiao2014a}, which support singlet-triplet mixing. 
Other studies also report unusual properties such as the suppression of spin-lattice relaxation rate coherence peak in NMR measurement \cite{matano_nmr_2013}, weak ferromagnetism near the transition temperature \cite{jha_weak_2015}, and topological band structure inferred from quantum oscillations \cite{Khan2019}. 
Furthermore, the presence of topological Dirac surface states have been reported by several photoemission studies \cite{thirupathaiah2016,Neupane2016,benia2016}, despite the fact that all experiments were conducted above the superconducting transition temperature, and that there are discrepancies in the interpretations of the observed band structure \cite{Yaresko2018}.

Presented the challenge in settling the nature of the superconducting state of $\alpha-$BiPd, it is desirable to perform not amplitude-sensitive, but phase-sensitive measurements of the pairing state \cite{tsuei_pairing_2000}. 
The single-valueness of the complex superconducting wave function demands a universal phase change of $2\pi$ in any closed path around a doubly-connected superconducting body, which leads to magnetic flux quantization \cite{byers_theoretical_1961}. 
First and routinely demonstrated in $s$-wave SCs, the fluxoid quantizes in interger numbers of flux quanta, or $\Phi'=n\Phi_0$, where $n$ is an integer number and $\Phi_0=hc/2e$ \cite{Deaver1961}. 
Anisotropic non-$s$-wave pairing, on the other hand, may induce an additional $\pi$ phase shift on crystalline grain boundaries \cite{geshkenbein_vortices_1987,Sigrist_Paramagnetic_1992}. 
Consequently, the flux quantization favors half-integer quanta, or $\Phi'=(n+1/2)\Phi_0$. 
As we have demonstrated in the case of centrosymmetric $\beta-$Bi$_2$Pd, anisotropic non-$s$-wave pairing symmetry can be unambiguously evidenced by half-quantum flux (HQF) quantization in polycrystalline ring devices \cite{Li_HQF_2019}. 
The distinctive experimental signature of HQF can be particularly powerful in determining the spin-triplet $p$-wave component in the presumed singlet-triplet mixture. 
In this work, we perform Little-Parks experiment \cite{little_observation_1962} to determine the magnetic flux quantization in polycrystalline rings of $\alpha-$BiPd. 
We report the observation of HQFs alongside with ordinary integer-flux quantization which provides a smoking-gun evidence for the presence of spin-triplet pairing and the pair mixing in noncentrosymmetric SCs. 

We prepared 50~nm-thick $\alpha-$BiPd thin films by magnetron sputtering, deposited onto SrTiO$_3$ (001) substrates held at elevated temperature of 400 $\rm ^oC$. 
The as-grown films are capped with 1~nm-thick MgO protecting layer before removing from the vacuum chamber. 
We obtain polycrystalline $\alpha-$BiPd films which are (1$\overline{1}$2)-textured [Fig.~1(c)]. The polycrystalline nature is revealed by the pole figure scan which shows no in-plane orientation (not shown). $\alpha-$BiPd films enter superconducting phase at the $T_c$ of 3.6~K with a sharp transition of less than 0.1~K, similar to those of bulk specimens \cite{joshi_superconductivity_2011}. 

The Little-Parks effect concerns the periodic oscillation of the free energy, and thus the oscillation of $T_c$, as a function of the applied magnetic flux threading through a superconducting ring \cite{little_observation_1962}. 
Experimentally the $T_c$ is measured by monitoring the electric resistance $R$ of the system, which is set at a fixed temperature within the transition regime just slightly below $T_c$. 
The experimental setup is depicted in Fig.~1(b). 
The typical Little-Parks effect for the well-known integer-flux quantization of $\Phi'=n\Phi_0$ is schematically presented in Fig.~1(d), where the resistance minima occur at the zero field and everywhere else when the applied flux $\Phi$ equals $n\Phi_0$. 
A $\pi$ phase shift induced by anisotropic gap function gives rise to the HQF, where the quantization condition becomes $\Phi'=(n+1/2)\Phi_0$ \cite{geshkenbein_vortices_1987}. 
This scenario would be expected for $\alpha-$BiPd if the pairing state indeed contains a spin-triplet component, as opposed to being overwhelmingly isotropic $s$-wave singlet pairing \cite{sun_dirac_2015,Peets2016}. 
As a result, the Little-Parks oscillation shifts a phase of $\pi$, or one half of a period, as shown in Fig.~1(e). 
In stark contrast, the resistance minima now occur when $|n-\Phi/\Phi_0|=\frac{1}{2}$. 
At zero field, the resistance reaches its maxima instead. 
To examine the Little-Parks effect, we fabricate the $\alpha-$BiPd thin films into various sub-$\micro$m-sized ring devices by electron beam lithography. 
The dimensions of the rings are chosen so that the oscillation period in term of magnetic field is reasonably large ($>$~20~Oe); therefore the zero-external-field state can be comfortably determined \cite{Li_HQF_2019}. 
A scanning electron microscopy (SEM) image of a representative $\alpha-$BiPd ring device is shown in Fig.~1(b). 
Typical for the sub-micron-sized devices, the superconductor-normal-metal transition broadens \cite{Li_HQF_2019} and the Little-Parks effect is typically observed in a temperature window between 2~K and 3~K, when the superconducting rings just start to become resistive. 

The Little-Parks effect distinctively reveals the presence of HQFs in polycrystalline $\alpha-$BiPd rings. 
In Fig.~2 we demonstrate such an example, observed in Device~A, a square-shaped ring which measures 450~nm between the opposing walls. 
The observed oscillation period of 106.2 Oe is in good agreement with the expected value of 102.1 Oe calculated from the enclosed area of the ring. 
The raw data, shown in Fig.~2(a) upper panel, demonstrates the oscillations on top of a roughly parabolic-shaped background, commonly observed in Little-Parks experiments \cite{little_observation_1962,Tinkham_1963,Moshchalkov_1995}. 
The background can be well described by a polynominal function (black dashed line) \cite{Li_HQF_2019}. 
One may subtract the background and obtain the oscillatory component $\Delta R$ versus $H$ as shown in the lower panel of Fig.~2(a). 
The resistance reaches maximum at the zero field as well as every integer numbers $\Phi_0$. 
It agrees with the scenario of HQF as depicted in Fig.~1(e). 

We take caution against artifacts such as trapped magnetic flux that could mimic HQF. 
Trapped flux may induce hysteresis behavior and shift an ordinary Little-Parks oscillation to the false appearance of a $\pi$ phase shift \cite{Li_HQF_2019}. 
The Little-Parks oscillation shown in Fig.~2(a) is symmetric with respect to the zero magnetic field, indicating the $\pi$ phase shift is not induced by defect-trapped vortices.
We may further rule out this potential artifact by sweeping the magnetic field in both directions, as shown in Fig.~2(b). 
Before each scan, the sample was first warmed up to 10~K, then cooled down in zero magnetic field. 
The opposite field scans yield virtually identical results, which shows no indication of trapped fluxes. 
The Little-Parks effect is also examined in an extended temperature range, as shown in Fig. 2(b), where the HQF remains robust in various temperatures. 

We have also observed HQFs in two other samples, in devices B and C. 
The results are summarized in Fig.~3. 
The $\pi$ phase shift can be observed for various temperatures and both field sweeping directions. 
The geometric factor varies among all three devices as well. 
Devices B (500~nm $\times$ 500~nm) and C (800~nm $\times$ 800~nm) are larger in size compared to Device A (450~nm $\times$ 450~nm). 
The wall widths for Devices A, B and C are 50~nm, 100~nm and 100~nm, respectively. 
We conclude that the pairing state of superconducting $\alpha-$BiPd must be anisotropic, which is necessary for supporting HQFs \cite{geshkenbein_vortices_1987}. 
This observation is consistent with the prediction of singlet-triplet pair mixing. 

In polycrystalline ring devices, the HQF rings (also known as the $\pi$-rings) are equally probable as the integer-quantum flux rings (the $0$-rings) \cite{Li_HQF_2019}. 
The realization of HQF is contingent upon the number of particular crystalline grain boundaries that produce a $\pi$ phase shift, or $\pi$-junctions \cite{geshkenbein_vortices_1987,Li_HQF_2019}. 
Only an odd number of $\pi$-junctions would produce a net phase change of $\pi$, which leads to a $\pi$-ring, whereas an even number of $\pi$-junctions where the total phase change adds up to 2$\pi$ leads to a $0$-ring. For a total of 16 $\alpha-$BiPd rings we measured, 3 $\pi$-rings and 13 $0$-rings are observed. 
In Fig.~4 we show three examples of $0$-rings, Devices A1, B1 and C1, which share the same design geometries with the three $\pi$-ring counterparts, respectively. 
They manifest the conventional integer-quantum flux quantization of $\Phi'=n\Phi_0$, as depicted in Fig.~1(d). 

It becomes interesting when the $\pi$-ring / $0$-ring ratio is put into context. 
For centrosymmetric $\beta-$Bi$_2$Pd, we found that more than 60\% of the total devices are $\pi$-rings. 
In stark contrast, $\pi$-rings of noncentrosymmetric $\alpha-$BiPd are conspicuously rare, or less than 20\% of the total devices. 
We find that the low $\pi/0$ ratio of $\alpha-$BiPd is indicative of the pair mixing nature of noncentrosymmetric SCs, assuming that the crystalline orientations of the grains are random. 
It is a common conclusion among the amplitude-sensitive studies \cite{sun_dirac_2015,joshi_superconductivity_2011,Peets2016} that the spin-singlet $s$-wave component dominates over the possible spin-triplet component. 
Our observation suggests that in noncentrosymmetric $\alpha-$BiPd there is a sizable isotropic $s$-wave component which makes forming $\pi$-rings much more difficult than centrosymmetric $\beta-$Bi$_2$Pd, where the pairing could be purely spin-triplet. 
At this point, it is not clear to us how the $\pi/0$ ratio may quantitatively gauge the pair mixing, which could be an intriguing topic for future studies. 

To conclude, we have observed HQFs in the noncentrosymmetric superconductor $\alpha-$BiPd, evidenced by the $\pi$ phase shift of the Little-Parks oscillations. 
Our finding unambiguously indicates the presence of the spin-triplet pairing component, consistent with the expectation of singlet-triplet pair mixing in noncentrosymmetric SCs. 
This result also supports the conclusions of topological band structure reported by the photoemission studies \cite{thirupathaiah2016,Neupane2016,benia2016} and quantum oscillations \cite{Khan2019}. 
With the presence of spin-triplet pairing, $\alpha-$BiPd is likely a topological superconductor, a suitable candidate material in which to search for Majorana fermions. 
Our method can be applied to other noncentrosymmetric SCs to determine the potential admixture of singlet and triplet pair states.

We gratefully acknowledge the support from the U.S. Department of Energy (DOE), Basic Energy Science award no. DESC0009390. Xiaoying Xu was supported in part by SHINES, an EFRC funded by U.S. DOE Basic Energy Science award no. SC0012670.

\end{text}

\clearpage

\begin{figure}
	\centering
	\includegraphics[width=16cm]{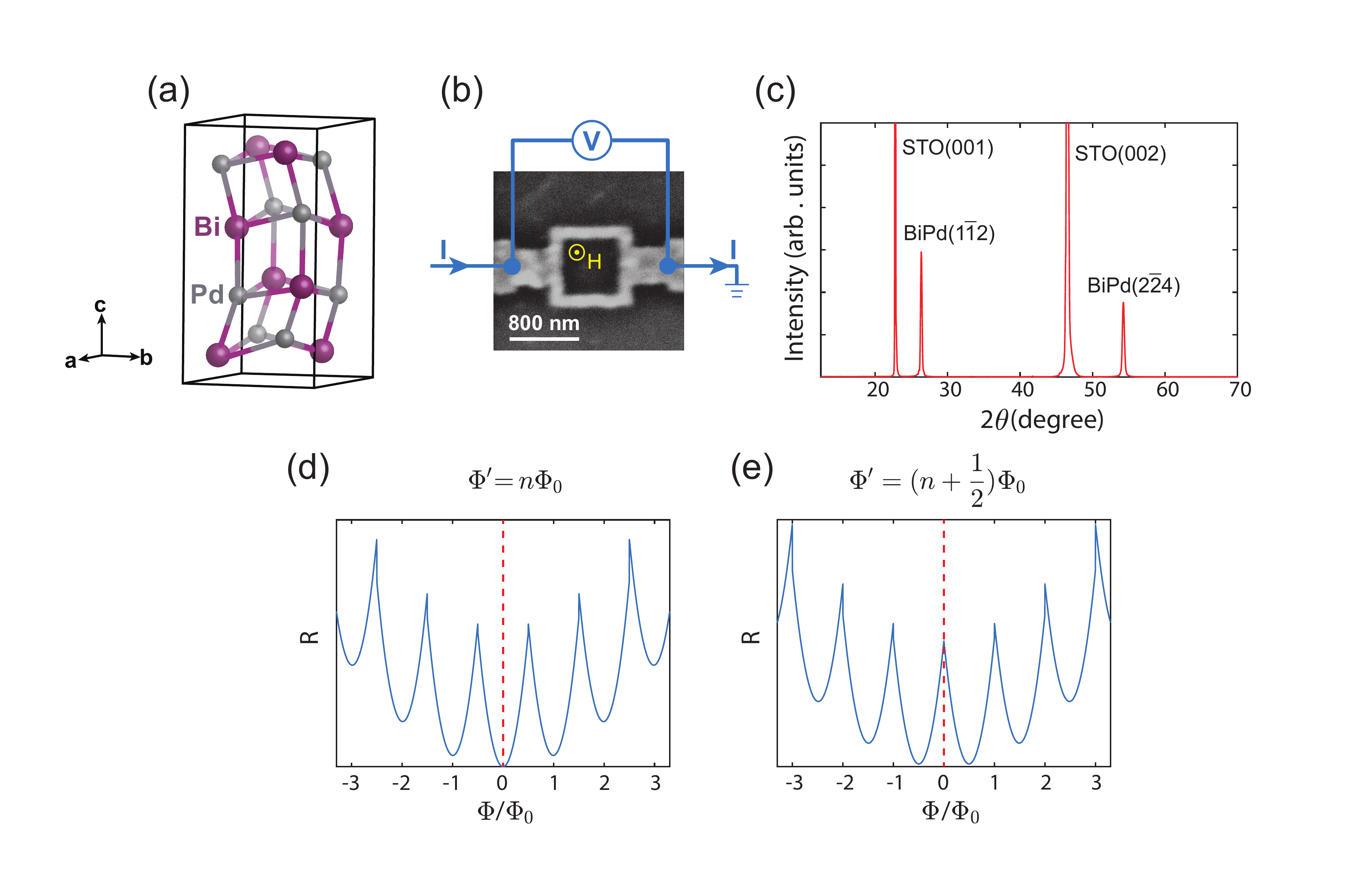}
	\caption{(a) Crystal structure of noncentrosymmetric superconductor $\alpha$-BiPd with space group $P2_1$. The lattice parameters are: $a=5.635 \mathrm{\AA}$, $b=5.661 \mathrm{\AA}$, $c=10.651 \mathrm{\AA}$, and $\gamma = 100.85\rm ^o$. 
	(b) The experimental setup of the ring structure with an out-of-plane magnetic field while the resistance is measured with a d.c. bias current of 1~$\micro$A. 
	The distance between the two opposing walls is 800~nm and the width of the side wall is 100~nm (Device C).   
	(c) X-ray diffraction spectrum of 50~nm-thick $\alpha-$BiPd thin film grown on SrTiO$_3$(001) substrate, 
	which shows the (1$\overline{1}$2)-textured plane of $\alpha-$BiPd parallel to the (001) plane of SrTiO$_3$. 
	Schematic drawing of the Little-Parks effect of a $0$-ring (d) with integer flux quantization: $\Phi^{'}=n\Phi_0$ and a $\pi$-ring (e) with half-integer flux quantization: $\Phi^{'}=(n+1/2)\Phi_0$.
	 } 
\end{figure}

\clearpage

\begin{figure}
	\centering    
	\includegraphics[width=16cm]{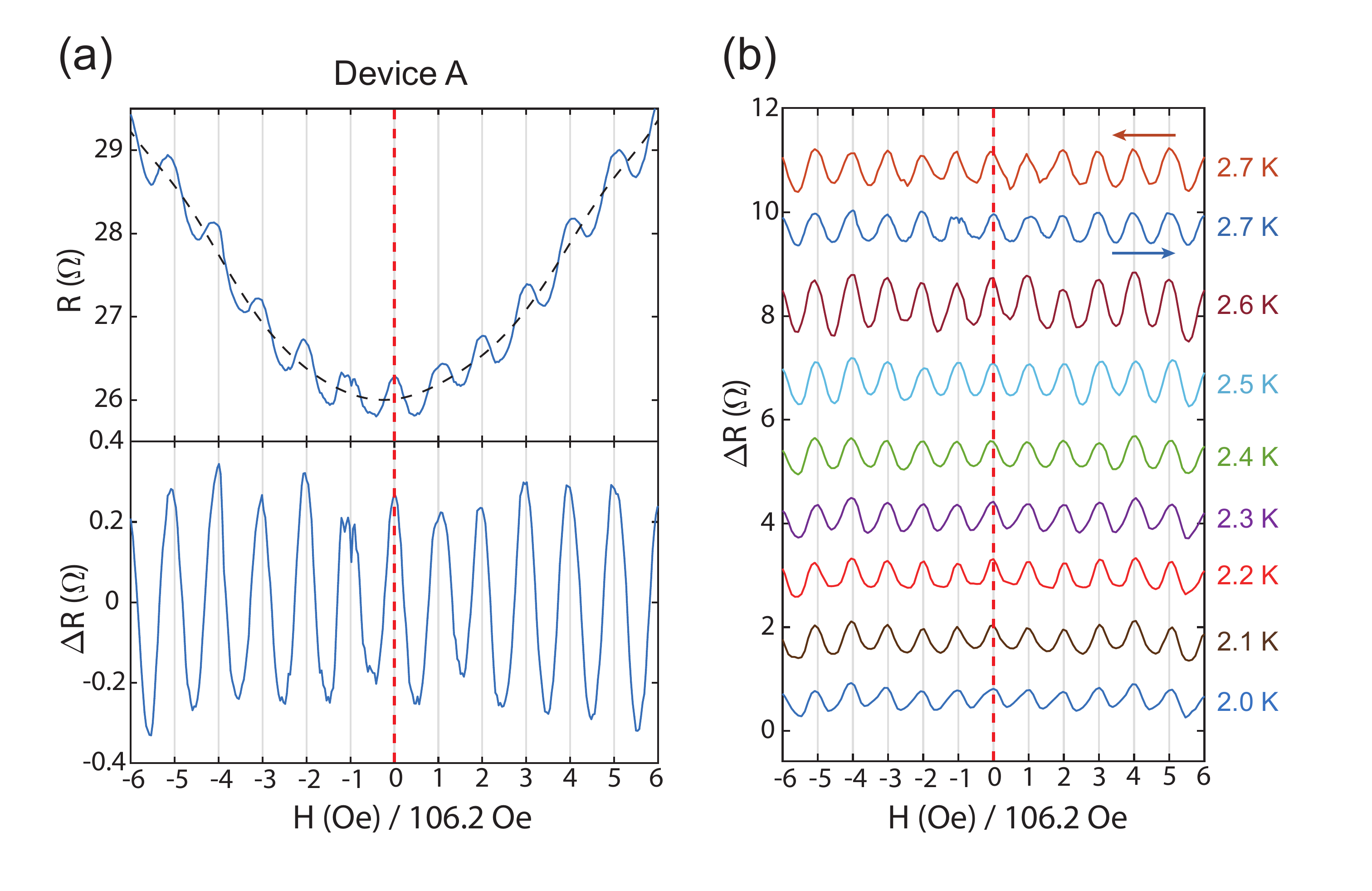}
	\caption{Little-Parks effect of Device A. (a) Upper panel: resistance as a function of applied magnetic field at 2.7 K. The red vertical dashed line denotes the zero field and the grey lines denote the fields at $n\Phi_0$. Device A has an enclosed area of 450 nm by 450 nm, which leads to an expected oscillation period of 102.1 Oe. The black dashed line is the fitted background curve. Lower panel: Little-Parks oscillation after subtraction of the background. (b) Temperature dependence of Little-Parks oscillations from 2 K to 2.7 K. The two curves at 2.7 K are obtained when sweeping the magnetic field in opposite directions. }
\end{figure}

\clearpage

\begin{figure}
	\centering
	\includegraphics[width=16cm]{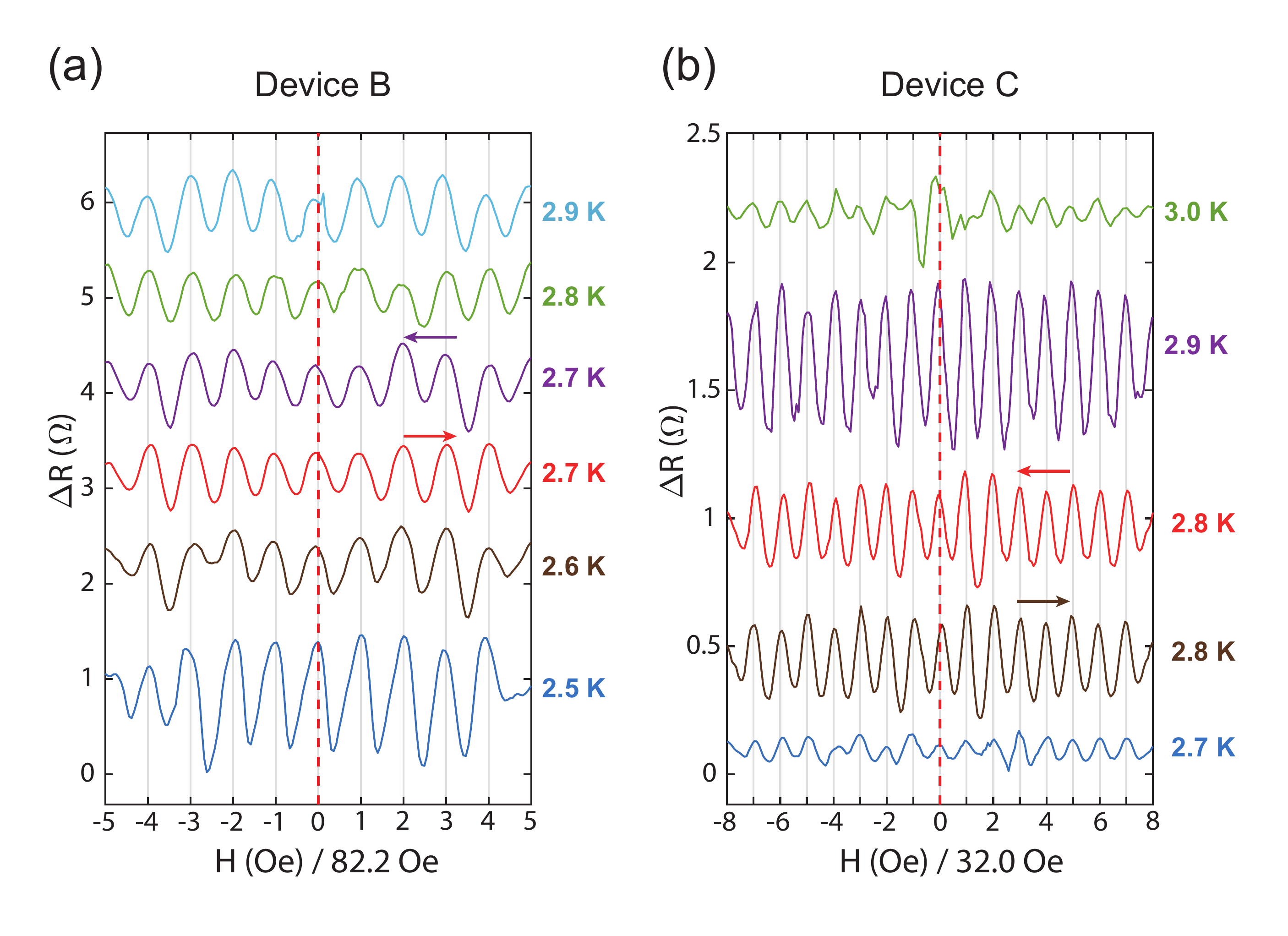}
	\caption{
		(a) Little-Parks effect of Device B (500~nm $\times$ 500~nm) at various temperatures with an expected oscillation period for $\Phi_0$ of 82.7~Oe. 
		(b) Little-Parks effect of Device C (800~nm $\times$ 800~nm) at various temperatures with an expected oscillation period for $\Phi_0$ of 32.3~Oe. 
		}
\end{figure}

\clearpage

\begin{figure}
	\centering
	\includegraphics[width=16cm]{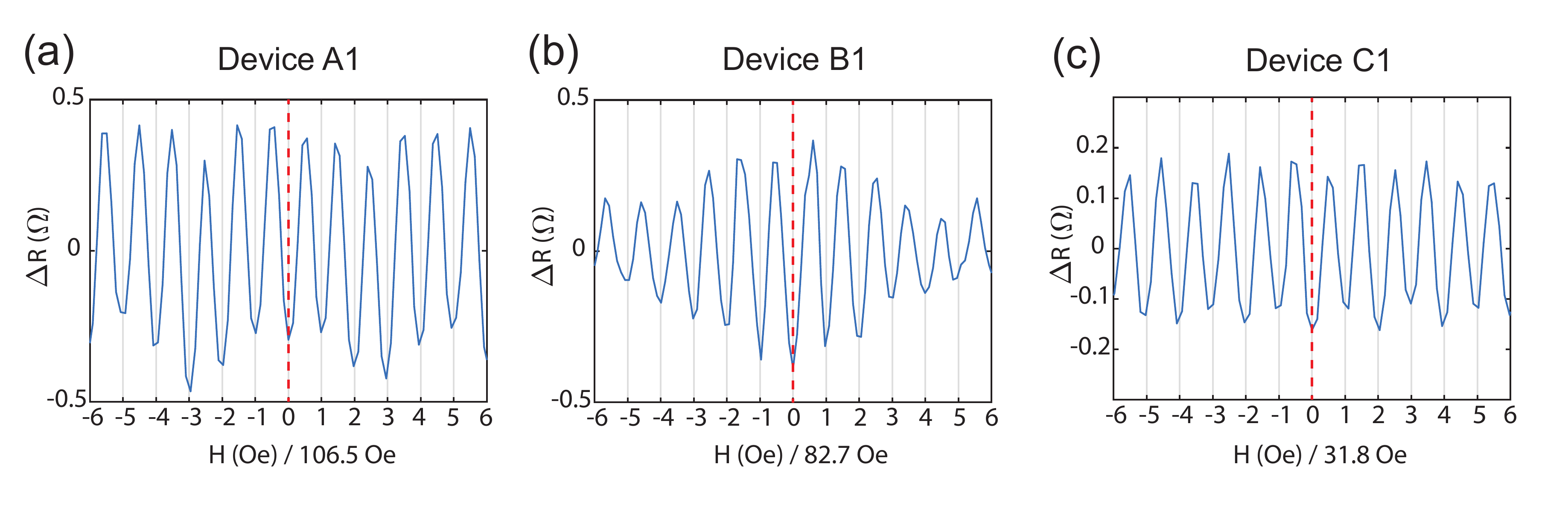}
	\caption{Little-Parks effect of three $0-$ring devices. 
		(a) Device~A1 (450~nm $\times$ 450~nm) at 2.6~K. 
		(b) Device~B1 (500~nm $\times$ 500~nm) at 2.9~K. 
		(c) Device~C1 (800~nm $\times$ 800~nm) at 2.9~K. }
\end{figure}

\end{document}